\documentstyle[12pt,titlepage,epsf]{article}

\newcommand{\be}{\begin{equation}}
\newcommand{\ee}{\end{equation}}
\newcommand{\bea}{\begin{eqnarray}}
\newcommand{\eea}{\end{eqnarray}}
\newcommand{\br}{\hskip .25cm/\hskip -.25cm}

\newcommand{\nonu}{\nonumber\\}
\newcommand{\la}{\hskip .25cm^\leftarrow\hskip -.25cm}

\newcommand{\ol}{\overline}

\begin{document}

\begin{center}
{\bf Functional Callan-Symanzik equation for QED}
\vspace{1cm}\\
J. Alexandre$^{1}$, J. Polonyi$^{2,3}$, and K. Sailer$^4$
\vspace{1cm}\\
$^1$ Department of Physics, King's College, London, UK\\
$^2$ Institute for Theoretical Physics, Louis Pasteur University,
Strasbourg, France\\
$^3$ Department of Atomic Physics, Lorand E\"otv\"os University
Budapest, Hungary\\
$^4$ Department for Theoretical Physics, University of Debrecen,
Debrecen, Hungary\\
\vspace*{1cm}
Abstract
\end{center}
An exact evolution equation, the functional generalization of the
Callan-Symanzik method, is given for the effective action of QED
where the electron mass is used to turn the quantum fluctuations
on gradually. The usual renormalization group equations are recovered
in the leading order but no Landau pole appears.

\vspace{2cm}

{\em Introduction:}
The idea of dimensional transmutation \cite{coleman} is that a dimensionless
parameter is traded for a dimensionful one. This replacement can be used
to generate a renormalization flow where the field amplitude which is a
quantity related to the size of the quantum fluctuations is evolved.
These flows ("fluctuation flows") prove to be equivalent with the usual
momentum flows at one-loop order. This scheme is developed in this paper for
QED by constructing the evolution of the one-particle irreducible
(1PI) generator functional where a mass parameter controls the
quantum fluctuations. We derive an exact equation describing this
evolution and will recover the usual one-loop momentum flows. We conclude
with some observations on the connection between the two renormalization
schemes.

The functional approach to renormalization group started
with the use of a 'sharp cut-off' $k$, i.e. the elimination of the
Fourier modes with $|p|>k$ was performed \cite{wh}. An infinitesimal
change $k\to k+\Delta k$ produces the renormalization group (RG) equation for
the Wilsonian, blocked action. This procedure has been
developed later \cite{whev}, as well as other schemes involving a 'smooth
cut-off', i.e. where the Fourier modes are suppressed
by means of a smooth, regulated version of the step-function
and the evolution of the effective action is sought \cite{polc,evol}.
The work presented here, although being quite different from a formal point
of view, is actually motivated by the smooth cut-off scheme
for the evolution of the effective action. The difference compared to that
method is that the suppression depends on the amplitude rather
than the characteristic wave-length of the modes.
This method has been developed for a scalar model \cite{jj} where
the usual RG equations were recovered in the U.V. regime,
i.e. where the mass is close to the cut-off. 
Finally, the same functional method was also used in QED with an external 
field \cite{ext} where the evolution of the generator functional 
of the one-particle irreducible (1PI) graphs 
with the amplitude of the external field led to the dependence of the
1PI graphs on the external gauge.

{\em Evolution equation:}
We start with the following bare Lagrangian in dimension $d=4-\varepsilon$
\be\label{barelag}
{\cal L}=\frac{1}{2e^2\mu^\varepsilon}{\cal A}_\mu\Box\left(T^{\mu\nu}+
\alpha L^{\mu\nu}\right){\cal A}_\nu
+\ol\Psi \left(i\br\partial-\br {\cal A}-zm_0\right)\Psi,
\ee
where $T^{\mu\nu}$ and $L^{\mu\nu}$ are respectively the
transverse and longitudinal projectors in the inverse photon propagator, and
the gauge parameter $\alpha$ characterizes the gauge fixing. The parameter
$z$ is introduced to control the amplitude of the fluctuations.
For $z>>1$ the theory is dominated by a free mass term contribution and
is perturbative. As $z$ decreases the interaction with the gauge
field becomes more important and quantum corrections increase in amplitude.
Our aim is to
study the evolution in $z$ of the generator functional $\Gamma_z$ of the 1PI
diagrams, the effective action.
The functional $W_z$ of the connected diagrams  is given by
\bea\label{cggen}
&&\exp W_z[\ol\eta,\eta,j^\mu]\\
&=&\int{\cal D}[\ol\Psi,\Psi,{\cal A}_\mu]
\exp\left\{i\int_x{\cal L}+i\int_x(j^\mu {\cal A}_\mu+\ol\eta\Psi+
\ol\Psi\eta)\right\}\nonumber,
\eea
and has the following functional derivatives
\bea\label{derivW}
\frac{\delta}{\delta\ol\eta}W&=&i\left<\Psi\right>=i\psi,\nonu
W\frac{\la\delta}{\delta\eta}&=&i\left<\ol\Psi\right>=i\ol\psi,\nonu
\frac{\delta}{\delta j_\mu}W&=&i\left<{\cal A}^\mu\right>=iA^\mu.
\eea
The effective action
$\Gamma_z[\ol\psi,\psi,A_\mu]$ is defined as the Legendre
transform of $W_z$,
\be
W_z=i\Gamma_z
+i\int_x\left(j^\mu A_\mu+\ol\eta\psi+\ol\psi\eta\right),
\ee
and has the following functional derivatives
\be\label{derivG}
\frac{\delta}{\delta\ol\psi}\Gamma=-\eta,
~~~~\Gamma\frac{\la\delta}{\delta\psi}=-\ol\eta,
~~~~\frac{\delta}{\delta A_\mu}\Gamma=-j^\mu .
\ee
The functional manipulations which lead to the evolution of $\Gamma_z$
are the same as those discussed in \cite{ext}: we first remark
that the derivatives (\ref{derivW}) and (\ref{derivG}) 
of the functionals $W_z$ and $\Gamma_z$ with 
respect to their variables imply that
\bea
\partial_z\Gamma&=&-i\partial_z W
=\left<-m_0\int_x\ol\Psi\Psi\right>\nonu
&=&-m_0\int_x\ol\psi\psi-m_0\int_x\frac{\delta}{\delta\ol\eta}
W\frac{\la \delta}{\delta\eta}.
\eea
Then the relation 
\be
\frac{\delta}{\delta\ol\psi}\Gamma\frac{\la\delta}{\delta\psi}=
-i\left(\frac{\delta}{\delta\ol\eta}W\frac{\la\delta}{\delta\eta}\right)^{-1}
\ee
leads us to the final equation for the  generator
functional of the 1PI graphs,
\be\label{evolG}
\partial_z \Gamma_z+m_0\int_x\ol\psi\psi=
im_0\mbox{Tr}\left\{
\left(\frac{\delta\Gamma\la\delta}{\delta\ol\psi\delta\psi}\right)^{-1}
\right\},
\ee
where 'Tr' denotes the trace over space-time and Dirac indices.
The inverse matrix $\delta^2\Gamma^{-1}$ has to be taken with respect to
the field variables, space-time and Dirac indices.
We stress that Eq.(\ref{evolG}) is exact and does not make
reference to any truncation of any sort.
Note the similarity between Eq.(\ref{evolG}) and other RG
 equations \cite{evol}. This similarity reflects the qualitative
agreement between the usual momentum flows and our method, as we will
discuss at the end of this paper.

{\em Gradient expansion:}
In order to find an approximate solution of the evolution equation,
we make an expansion in the amplitude and the gradient of the
fields and use the ansatz
\bea\label{gradexp}
\Gamma_z[\ol\psi,\psi,A_\mu]&=&
\int_x\biggl[A_\mu(x){\cal D}^{-1\mu\nu}A_\nu(x)\nonu
&&+\ol\psi(x){\cal G}^{-1}\psi(x)-Z(z)\ol\psi(x)\br A(x)\psi(x)\biggr],
\eea
where the photon and fermion propagators are
\bea
{\cal D}^{-1\mu\nu}(p)&=&\frac{\beta_T(z)}{e^2\mu^\varepsilon}p^2
T^{\mu\nu}(p)+
\alpha\frac{\beta_L(z)}{e^2\mu^\varepsilon}p^2L^{\mu\nu}(p),\nonu
{\cal G}^{-1}(p)&=&Z(z)\br p-m(z)-zm_0,
\eea
with $T^{\mu\nu}(p)=g^{\mu\nu}-p^\mu p^\nu/p^2$ and
$L^{\mu\nu}(p)=p^\mu p^\nu/p^2$.
The evolution of $\beta_T(z)$, $\beta_L(z)$, $Z(z)$ and $m(z)$ are
obtained by expanding both sides of Eq.(\ref{evolG}) in
powers of the gauge field and
by identifying the operators on the left and right hand sides.
For this end we need the inverse of the matrix $\Gamma^{(2)}$
\be
\Gamma^{(2)}=
\pmatrix{\frac{\delta\Gamma\la\delta}{\delta\ol\psi\delta\psi}&
\frac{\delta^2\Gamma}{\delta\ol\psi\delta\ol\psi}&
\frac{\delta^2\Gamma}{\delta A_\mu\delta\ol\psi}\cr
\frac{\Gamma\la\delta^2}{\delta\psi\delta\psi}&
\frac{\delta\Gamma\la\delta}{\delta\psi\delta\ol\psi}&
\frac{\delta\Gamma\la\delta}{\delta A_\mu\delta\psi}\cr
\frac{\delta\Gamma\la\delta}{\delta\psi\delta A_\mu}&
\frac{\delta^2\Gamma}{\delta\ol\psi\delta A_\mu}&
\frac{\delta^2\Gamma}{\delta A_\mu\delta A_\nu}},
\ee
which will be obtained by expanding in the diagonal part $\Gamma^{(2)}_\Delta$
in momentum space,
\bea
\left(\Gamma^{(2)}\right)^{-1}&=&
\left(\Gamma^{(2)}_\Delta\right)^{-1}-\left(\Gamma^{(2)}_\Delta\right)^{-1}
\Gamma^{(2)}_{nd}\left(\Gamma^{(2)}_\Delta\right)^{-1}\\ &+&
\left(\Gamma^{(2)}_\Delta\right)^{-1}
\Gamma^{(2)}_{nd}\left(\Gamma^{(2)}_\Delta\right)^{-1}
\Gamma^{(2)}_{nd}\left(\Gamma^{(2)}_\Delta\right)^{-1}\nonu &-&
\left(\Gamma^{(2)}_\Delta\right)^{-1}
\Gamma^{(2)}_{nd}\left(\Gamma^{(2)}_\Delta\right)^{-1}
\Gamma^{(2)}_{nd}\left(\Gamma^{(2)}_\Delta\right)^{-1}
\Gamma^{(2)}_{nd}\left(\Gamma^{(2)}_\Delta\right)^{-1}+...,\nonumber
\eea
where $\Gamma^{(2)}_{nd}=\Gamma^{(2)}-\Gamma^{(2)}_\Delta$.
We obtain in this manner the result
\bea
\left(\frac{\delta\Gamma\la\delta}{\delta\ol\psi\delta\psi}\right)^{-1}&=&
{\cal G}+{\cal G}\Big(-c+c{\cal G}c-c{\cal G}c{\cal G}c
+a^\mu{\cal D}_{\mu\nu}\tilde b^\nu\nonu
&&-a^\mu{\cal D}_{\mu\nu}\tilde b^\nu
{\cal G}c-c{\cal G}a^\mu{\cal D}_{\mu\nu}\tilde b^\nu\Big){\cal G}+...,
\eea
where
\bea
a^\mu(p,q)&=&-Z(z)\gamma^\mu\psi(-p-q),\nonu
b^\nu(p,q)&=&\ol\psi(-p-q)Z(z)\gamma^\nu,\nonu
c(p,q)&=&-Z(z)\br A(-p-q),
\eea
and the tilde stands for the transposed in momentum and spinor spaces.
The evolution equations read finally as
\bea\label{evolprops}
&&\partial_z {\cal D}^{-1\mu\nu}(p)=
-2im_0Z^2(z)\int_q tr\left\{{\cal G}^2(q)\gamma^\mu{\cal G}(p+q)\gamma^\nu
\right\}, \nonu
&&\partial_z{\cal G}^{-1}(p)+m_0=
-im_0Z^2(z)\int_q {\cal D}_{\mu\nu}(q-p)\gamma^\mu
{\cal G}^2(q)\gamma^\nu,\nonu
&&\partial_z Z(z)\gamma^\mu=
-2m_0Z^3(z)\int_q{\cal D}_{\kappa\nu}(q)\gamma^\nu
{\cal G}^2(q)\gamma^\mu{\cal G}(q)\gamma^\kappa.
\eea

The following remarks are important:
\begin{itemize}
\item The photon does not acquire a mass in the evolution in $z$ and
the IR divergences are the usual ones.
\item The longitudinal part of the photon propagator does not evolve
with $z$. Accordingly, the fluctuations do not
generate longitudinal contributions to photons.
\end{itemize}
Furthermore a technical point, it is the evolution equation of the
fermion mass only which needs regulator, the other equations are
divergence-free with the given effective action ansatz.

To compare this result with the traditional method based on the loop-expansion,
note that the change $zm_0\to(z+\delta z)m_0$
of the bare electron mass changes the fermion propagator as
${\cal G}\to{\cal G}+{\cal G}\delta zm_0{\cal G}$ in the internal lines
of the Feynman graphs. This observation, the starting point of the
Callan-Symanzik method is sufficient to realize that the evolution
equations (\ref{evolprops}) can obviously be written as
\bea\label{oneloopevolprops}
&&\partial_z {\cal D}^{-1\mu\nu}(p)=
-i\partial_z\int_q tr\left\{{\cal G}(q)\gamma^\mu{\cal G}(p+q)\gamma^\nu
\right\}+{\cal O}(\hbar^2) ,\nonu
&&\partial_z{\cal G}^{-1}(p)+m_0=
-i\partial_z\int_q {\cal D}_{\mu\nu}(q-p)\gamma^\mu
{\cal G}(q)\gamma^\nu+{\cal O}(\hbar^2),\nonu
&&\partial_z Z\gamma^\mu=
-\partial_z\int_q{\cal D}_{\kappa\nu}(q)\gamma^\nu
{\cal G}(q)\gamma^\mu{\cal G}(q)\gamma^\kappa+{\cal O}(\hbar^2),
\eea
whose integrals give immediately
\bea\label{oneloopgraphs}
&&{\cal D}^{-1\mu\nu}(p)={\cal D}^{-1\mu\nu}_{tree}(p)
-i\int_q tr\left\{{\cal G}(q)\gamma^\mu{\cal G}(p+q)\gamma^\nu\right\}
+{\cal O}(\hbar^2),\nonu
&&{\cal G}^{-1}(p)={\cal G}^{-1}_{tree}(p)
-i\int_q {\cal D}_{\mu\nu}(q-p)\gamma^\mu{\cal G}(q)\gamma^\nu
+{\cal O}(\hbar^2),\nonu
&&Z\gamma^\mu=Z_{tree}\gamma^\mu
-\int_q{\cal D}_{\kappa\nu}(q)\gamma^\nu{\cal G}(q)\gamma^\mu{\cal G}(q)
\gamma^\kappa+{\cal O}(\hbar^2),
\eea
the usual one-loop corrections to the 1PI functions appearing in the ansatz
(\ref{gradexp}).

{\em Beta functions:}
The computation of the integrals is straightforward and leads to the following
result,
\bea\label{seteq}
\beta_L'(z)&=&0,\nonu
\beta_T'(z)&=&\frac{e^2\mu^\varepsilon}{6\pi^2}\frac{m_0}{zm_0+m(z)},\nonu
Z'(z)&=&\frac{e^2\mu^\varepsilon}{8\pi^2\alpha}\frac{Z(z)m_0}{zm_0+m(z)},\nonu
m'(z)&=&-\frac{m_0}{8\pi^2}\frac{e^2\mu^\varepsilon}{\varepsilon}
\left(\frac{3}{\beta_T(z)}+\frac{1}{\alpha}\right)+\mbox{finite},
\eea
where the prime denotes a derivative with respect to $z$.
In what follows, we will keep the bare theory and $\varepsilon\ne 0$. We
define $m(z)=m_0\phi(z)$ and write $\beta(z)=\beta_T(z)$, such that
the equations (\ref{seteq}) read 
\bea\label{finiteequadiff}
\phi'(z)&=&-\frac{e^2\mu^\varepsilon}{8\pi^2\varepsilon}
\left(\frac{3}{\beta(z)}+\frac{1}{\alpha}\right),\nonu
\beta'(z)&=&\frac{e^2\mu^\varepsilon}{6\pi^2}\frac{1}{z+\phi(z)},\nonu
Z'(z)&=&\frac{e^2\mu^\varepsilon}{8\pi^2\alpha}\frac{Z(z)}{z+\phi(z)}.
\eea

In order to compare these scaling laws with the one-loop result,
we look at the situation of weak fluctuations where the fermion 
mass term is dominant in the bare action, i.e. where $z$ is close to 
some initial value $z_0>>1$ and the parameter $m(z)$ is close
to zero. In this regime we have $z+\phi(z)\simeq z$ and
$Z(z)\simeq\beta(z)\simeq 1$.
This approximation is equivalent to keeping in Eqs.(\ref{finiteequadiff}) 
the terms of the order $e^2$ only:
\bea\label{perturb}
\phi'(z)&\simeq&-\frac{e^2\mu^\varepsilon}{2\pi^2\varepsilon},\nonu
z\beta'(z)&\simeq&\frac{e^2\mu^\varepsilon}{6\pi^2},\nonu
z Z'(z)&\simeq&\frac{e^2\mu^\varepsilon}{8\pi^2}
\eea
in the Feynman gauge ($\alpha=1$). 
We see that if we make the identification $z/z_0=\mu_0/\mu$,
where $\mu_0$ is some IR scale smaller than $\mu$, we find the well-known
one-loop results \cite{pokorski}, since we 
have then $z\partial_z=-\mu\partial_\mu$ (a decrease of $z$ starting from $z_0$ is 
equivalent to an increase
of $\mu$ starting from $\mu_0$). One can understand
this by noting that the decrease $zm_0\to(z+\delta z)m_0$ of the
electron mass influences mainly the propagation with momentum
$p\approx zm_0$. In such a manner the modes whose dynamics is
included in the effective action as $z$ decreases ($dz<0)$
are approximately those which are included in the traditional
RG step when the momentum scale of the cut-off is increased
from $p=(z+\delta z)m_0$ to $p=zm_0$.
This picture is lost for 1PI graphs containing two
or more loops since the momenta of the internal loops appear as another scale
which modifies the effect of the changing regulator parameters
in a more complicated way. In this case
our scheme will be different from the usual one, reflecting the 
different organization of the gradient and loop 
expansions.

{\em Landau pole:}
It is interesting to note that the integration of equation (\ref{perturb}),
\be
\beta(z)\simeq 1+\frac{e^2}{6\pi^2}\ln(z/z_0)
\ee
reproduces the well-known one-loop value of the Landau-pole,
\be\label{landaupole}
\frac{z}{z_0}=\frac{\mu_0}{\mu}=\exp\left(-\frac{6\pi^2}{e^2}\right).
\ee
But the essential difference is that while the traditional Landau-pole
is obstructing the further increase of the UV cut-off, the pole
in the present scheme prevents us to further decrease the bare mass
and indicates some problem with large amplitude quantum fluctuations.
Since the initial condition of our evolution equation is already an effective
potential corresponding to the regulated theory with a fixed cut-off, 
the UV Landau pole cannot be seen in our scheme. In both cases the pole
reflects the problem that the loop corrections dominate the tree-level part
of the effective action when 'too many' modes are treated in the one-loop
approximations.

But this similarity is an artifact of the autonomous solution of the 
truncated equation (\ref{perturb}). Instead one should allow the other 
parameters, at least the fermion mass, to run as well. For this end we write the
evolution equation of $\beta$ as
\be
\frac{1}{\beta'(z)}=\frac{6\pi^2}{e^2\mu^\varepsilon}(z+\phi(z)),
\ee
which gives
\be
\frac{\beta''(z)}{\beta'(z)}=\left(\frac{3}{4\alpha\varepsilon}-
\frac{6\pi^2}{e^2\mu^\varepsilon}\right)\beta'(z)+\frac{9}{4\varepsilon}
\frac{\beta'(z)}{\beta(z)},
\ee
after using the first equation of (\ref{finiteequadiff}). It is easy to integrate 
this equation and find by means of (\ref{finiteequadiff}) again:
\be\label{nolandaupole}
\beta(z)=\left(\frac{z_0+\phi(z_0)}{z+\phi(z)}\right)^{4\varepsilon/9}
\exp\left\{\frac{1-\beta(z)}{3}\left(\frac{1}{\alpha}
-\frac{8\pi^2\varepsilon}{e^2\mu^\varepsilon}\right)\right\}
\ee
It should be noted that we cannot na\"{\i}vely take the limit $\varepsilon\to 0$
in the expression (\ref{nolandaupole}) since $\phi(z)$ is then diverging.
This equation shows that $\beta(z)$ is non-vanishing for any value of $z$ and therefore 
this pole is actually avoided by the evolution of the fermion mass
and there is no problem to reconstruct the physical theory at $z=0$.
It is remarkable that the mechanism by which this happens, a running fermion mass,
is actually the same how the Landau pole is claimed to be avoided in 
lattice QED by spontaneous symmetry breakdown of the chiral symmetry
\cite{gockler}.

{\em Conclusion:}
To conclude, we stress the analogies between the momentum dependence
traced by the usual renormalization procedures and the dependence on the
amplitude of quantum fluctuations. We showed that the leading order
scaling law corresponding to small amplitudes (i) agrees with the traditional
scaling law of the U.V. regime and (ii) is independent of the choice of
the gauge as shown by Eqs.(\ref{oneloopgraphs}). The present work is
the first step towards
an extension of the renormalization scheme based on the field
amplitude for  non-Abelian gauge theories where a suitable
gradient expansion should provide us with a deeper insight
into the dynamics of the I.R. regime.

\vspace{1cm}
{\bf Acknowledgement:} This work has been supported in parts by the
TMR E.U. contract FMRX-CT97-0122, the Leverhulme Trust, U.K.,
the grants OTKA T29927/98, OTKA T032501/00, and  DFG-MTA 436UNG113/140/0.


\begin{thebibliography}{99}
\bibitem{coleman} S.Coleman, E.Weinberg, Phys.Rev.D7: 1888 (1973).
\bibitem{wh} F.J.Wegner, A.Houghton, Phys.Rev A8: 401 (1973).
\bibitem{whev} J.M.Carmona, J.Polonyi, A.Tarancon, Phys.Rev.D61:
085018 (2000); J.Alexandre, V.Branchina, J.Polonyi, Phys.Lett.B445: 351 (1999);
S.B.Liao, J.Polonyi, Phys.Rev.D51: 4474(1995).
\bibitem{polc} J.Polchinski, Nucl.Phys.B231: 269 (1984).
\bibitem{evol} C.Wetterich, Phys.Lett.B301: 90 (1993);
M.Reuter, C.Wetterich, Nucl.Phys.B391, 147 (1993);
T.Morris, Int.J.Mod.Phys.A9: 2411 (1994);
U.Ellwanger, Phys.Lett.B335, 364 (1994);
J.Bergers, N.Tetradis, C.Wetterich, hep-ph/0005122;
S.Arnone, Yu.A.Kubyshin T.R.Morris, K.Yoshida, J.Mod.Phys.A16: 1809 (2001);
U.Ellwanger, Z.Phys.C76: 721 (1997);
M. Simionato, Int. J. Math. Phys. A15:2121(2000); 2153 (2000).
\bibitem{jj} J. Alexandre, J. Polonyi, Ann.Phys.286: 1 (2000).
\bibitem{ext} J. Alexandre, Phys.Rev.D64: 045011 (2001).
\bibitem{pokorski} S.Pokorski, {\it Gauge Field Theories}, Cambridge
Monographs on Mathematical Physics (1987), section 5-2.
\bibitem{gockler} M.G\"ockler, R.Horsley, V.Linke, P.Rackow, G.Shierholtz,
H.St\"uben, Phys.Rev.Lett.80: 4119 (1998).
\end{thebibliography}
\end{document}